\documentstyle[11pt]{article}
\def \be{\begin{equation}}
\def \ee{\end{equation}}
\normalbaselines
\begin{document}
\begin{flushright}
TIFR/TH/93-50
\end{flushright}
\bibliographystyle{unsrt}
\begin{center}
{\LARGE \bf Effects of Possible $\Delta B =- \Delta Q$ Transitions in
Neutral $B$ Meson Decays}\\[5mm]
G. V. Dass$^1$ and K. V. L. Sarma$^2$\\[3mm]
$^1${\it Department of Physics, Indian Institute of Technology,
                   Powai, Bombay, 400 076, India    }\\
$^2${\it Tata Institute of Fundamental Research, Homi Bhabha
Road, Bombay, 400 005, India\\
(e-mail: kvls@tifrvax.bitnet)}\\[10mm]
\end{center}
\begin{quotation}
\small{ We explore the possibility that the existing data on
like-sign dileptons at the $\Upsilon (4S)$ resonance consist of events
arising from $B_{d}^0 -\bar B_{d}^0$ mixing and also from
$\Delta B = - \Delta Q$ transitions. The consequences of these nonstandard
transitions for certain time-asymmetries which are likely to be measured at
the $B$ factories are studied.
\bigskip

  PACS numbers: 13.20.-v, 12.15.Ji.}
  \end{quotation}
\baselineskip=0.8cm

     An interesting way in which physics beyond the Standard Model could
manifest itself is through violations of the $\Delta B = \Delta Q$ rule in
the decays of bottom hadrons. These violations if present in decays of
neutral $B$ mesons will have obvious implications for the study of $B\bar
B$ mixing because experimentally the bottom flavour is tagged by the decay
leptons. It is therefore necessary to examine the effect of possible
$\Delta B = -\Delta Q$ transitions on the same-sign dilepton signal which
is observed at the $\Upsilon(4S)$ resonance \cite{ARG87,CLE89,BAR93}. This
is because not all the `wrong-sign' leptons might be originating from
$\Delta B = 2$ transitions, $B\leftrightarrow \bar B$; some may be coming
from decays that do not obey the $\Delta B = \Delta Q$ rule, \cite{FN1}.
In terms of quarks this amounts to postulating the transition
$b\rightarrow \beta~+ ~W^+$, where $\beta$ is an exotic quark carrying a
charge of (${-4/3}$) units. Such exotic quarks have been envisaged in
certain models; for a recent model see, e.g., Ref.\cite{FOO93}. These
quarks could lead to the existence of exotic mesons which are doubly
charged (e.g., $\beta \bar u$), or to a spectacular jump in the $e^+e^-$
annihilation ratio $R$ by (16/3) units. Available data upto LEP energies
do not indicate any evidence for such quarks. Also the present data on the
semileptonic decays of neutral beons neither require nor forbid $\Delta B
= -\Delta Q$ transitions which lie outside the Standard Model.

                Recently Kobayashi and Sanda \cite{KS92} have suggested
some tests for checking $CPT$ invariance at the $B$ factories. They
assumed the $\Delta B=\Delta Q$ rule of the Standard Model as being valid
in semileptonic decays. On the other hand they relaxed $CPT$ invariance
for the mass-matrix of the neutral beons, but not for decay amplitudes. In
the following we take a complementary approach: We assume the validity of
complete $CPT$ invariance throughout, but allow $\Delta B= -\Delta Q$
transitions. We examine how the $\Delta B=-\Delta Q$ amplitudes could
generally affect the dilepton ratios and certain time-asymmetries that are
measurable at the $B$ factories. Of particular interest in this context
are the time-integrated asymmetries in dilepton events which arise from
exclusive semileptonic decays of neutral $B$'s.  We further show that the
asymmetries which involve the detection of $CP$ eigenstates in the decays
of $B$ and $\bar B$, namely, the well-known $CP$-violating asymmetries
\cite{CS80,BI89,NQ92} and the $y$-determining asymmetry \cite{DS92} are
unlikely to be affected much by the $\Delta B=-\Delta Q$ transitions.

   {\it Notation~-~} For notational convenience in the following, we drop
the superscripts and subscripts on the mesons $B_d^0 $ and $\bar B_d^0$,
and refer to them simply as $B$ and $\bar B$, respectively. Mixing allows
the construction of eigenstates which have definite masses $m_{1,2}$ and
inverse lifetimes $\Gamma _{1,2}$ in the usual way,
\[ |B_1> =p|B>+~q|\bar B>~,~~|B_2>=p|B>-~q|\bar B>~,~~~(|p|^2 +|q|^2=1)~;
\]
they evolve in proper time $t$ as \cite{Kh90}
\[  |B_k>~\rightarrow ~e^{(-im_k-{1 \over 2}\Gamma _k)t}~|B_k>~;~~(k=
1,2)~.  \]
We assume $CPT$ invariance throughout, and take $|\bar B>~=~CP|B>$ . We
define the symbols
 \[ g~\equiv~{q\over p}~,~~x~=~{m_2~-~m_1\over \Gamma }~,~~ y~=~{
 \Gamma _2~-~\Gamma _1 \over {2~\Gamma }}~,~~ \Gamma ~=~{\Gamma _1+
 \Gamma _2\over 2}~.\]

    For notational brevity we refer to a particular channel of
semileptonic beon decay merely by the corresponding hadronic state label,
and distinguish its $CPT$-conjugate channel by a `tilde':
\[ (~i\ell ^+)\equiv (X_i + \nu_\ell +\ell^+)~~,~~~(~\tilde i\ell^-)
     \equiv (\bar X_i + \bar \nu_\ell~ + \ell^-)~~;\]
thus the labels $i$ or $\tilde i$ are taken to include the appropriate
neutrinos. The decay amplitudes obeying the $\Delta B = \Delta Q$ rule
will be denoted by
\be  A_i~= ~ <i \ell ^+|~ T~ | B>~,~~ \bar A_i~ =~
                             < \tilde i \ell ^-|~ T~ |\bar B>~;  \ee
the corresponding amplitudes for $\Delta B = -\Delta Q$ transitions will
contain complex multiplicative parameters $\rho$ as follows:
  \be  \rho _i A_i~ =~ <i\ell ^+|~ T~ |\bar B>~,~~
       \bar \rho _i \bar A_i ~=~ <\tilde i \ell ^-|~ T~ |B>.\ee
Neglecting the final-state interactions due to electroweak forces, $CPT$
invariance leads to the relations
  \be \bar A_i ~=~ A_i^*~,~~\bar \rho _i ~=~ \rho _i^*~.\ee
In the Standard Model, the $\rho $'s are expected to be small since they
get contributions from diagrams involving at least two $W$'s; indeed in the
case of kaons the ratio of the $\Delta S=-\Delta Q$ and $\Delta S=\Delta Q$
amplitudes is estimated \cite{DIB91} to have a magnitude of order
$10^{-7}$. In what follows we shall regard the $\rho $'s as small
parameters.

 {\it Ratio of Dilepton Events~-~} We consider exclusive semileptonic
decays of the neutral $B$ mesons into the channels $(i \ell ^+)$ and
$(j \ell ^+)$, where, for instance, $i$ and $j$ could stand for the states
$(D^{*-}\nu _\ell )$ and $(D^-\nu _\ell )$. The numbers of
events with same-sign dileptons from the $\Upsilon (4S)$, irrespective of
the decay times, are given (apart from an overall constant) by
 \begin{eqnarray}
 n(i\ell ^+,j\ell ^+)&=&C_{ij}\left \{(1-a)|1-r_ir_j|^2
                               +(1+a)|r_i-r_j|^2\right \}~,\\
  n(\tilde i\ell ^-,\tilde j\ell ^-)&=&C_{ij}~|g|^4~\left \{(1-a)
         |1- \bar r_i\bar r_j|^2 ~+~(1+a)|\bar r_i-\bar r_j|^2\right \}~;
    \end{eqnarray}
 \be C_{ij}~\equiv ~
 \left|{A_iA_j \over {\sqrt{2} g\Gamma }}\right|^2{1\over {1-y^2}}~~,
                           ~~~a={{1-y^2} \over {1+x^2}}~~, \ee
 \be   r_i= g \rho _i~, ~~ \bar r_i~=~(\bar \rho _i /g)~. \label{orr}\ee
 Similarly, the (relative) numbers of events with opposite-sign dileptons,
integrated over all times, are given by
 \[ n(i\ell ^+,\tilde j\ell ^-)~=~C_{ij}~|g|^2\left \{(1+a)|1-r_i\bar r_j|
 ^2  +(1-a)|r_i-\bar r_j|^2\right \}~,\]
 \[ n(j \ell^+,\tilde i\ell^-)~=~C_{ij}~|g|^2\left \{(1+a)|1-r_i
 \bar r_j|^2  +(1-a)|r_j-\bar r_i|^2\right \}~,\]
where we used $|1-r_i\bar r_j|=|1-\bar r_ir_j|$.

                We next define the `exclusive' dilepton ratio
$\chi _{ij}$, as the number of like-sign dilepton events relative to the
total number, where all the events originate in either of the two
exclusive decays having labels $i$ and $j$, or their conjugates
$\tilde i$ and $\tilde j$:
  \be      \chi _{ij}~=~ {N_{ij}^{++}~+~N_{ij}^{--} \over
       N_{ij}^{++} ~+~N_{ij}^{--} ~+~N_{ij}^{+-}~+~N_{ij}^{-+}}~,\ee
 \begin{eqnarray} N_{ij}^{++} &\equiv & n(i\ell ^+,i\ell ^+)+
 n(i\ell ^+,j\ell ^+) +n(j\ell ^+,j\ell ^+)~,   \nonumber\\
     N_{ij}^{--}  &\equiv &  n(\tilde i\ell ^-,\tilde i\ell^-)+
 n(\tilde i\ell ^-,\tilde j\ell ^-)
 +n(\tilde j\ell ^-,\tilde j\ell ^-)~,\nonumber\\
 N_{ij}^{+-}+N_{ij}^{-+} &\equiv &   n(i \ell ^+,\tilde i\ell ^-)+
 n(i \ell ^+,\tilde j\ell ^-)+   n(j \ell ^+,\tilde i\ell^-)+
 n(j\ell ^+,\tilde j \ell ^-)~. \nonumber  \end{eqnarray}

   In the following we treat $\rho$ and $\bar \rho$ to be small in
magnitude and keep terms upto and including ${\it second~order}$ in them;
for example, we shall write
\[ |1-r_i\bar r_j|^2=|1-(r_ir_j^*/|g|^2)|^2 \simeq [1-2Re~(r_ir_j^*)]~,\]
where in the last step we ignored the additional correction due to $CP$
violation by setting $|g|=1$ as it is multiplying the quadratically
small quantity $(r_ir_j^*)$.

    Thus the modified ratio of like-sign dilepton events arising from
either of the two channels $i$ and $j$ is
   \be      \chi _{ij}= \left\{ 1~+~4a~(Im~<r>_{ij})^2~+~
     {4a \over 1-a}\left({|A_iA_j|~|r_i -r_j| \over |A_i|^2+|A_j|^2}
                            \right)^2\right\}~ \chi~.\ee
Here $\chi$ is the usual dilepton ratio for inclusive channels
assuming the $\Delta B=\Delta Q$ rule \cite{OPZ75},
\be \chi~ =~ {(1-a)(1+|g|^4) \over (1-a)(1+|g|^4)~+~2(1+a)|g|^2}~;\ee
$Im~<r>_{ij}$ is the imaginary part of the weighted average of the ratios
$r_i$ and $r_j$ ,
\be <r>_{ij}~\equiv ~{r_i~|A_i|^2 +r_j~|A_j|^2 \over |A_i|^2+|A_j|^2}~~,
  ~~r_i~=~{q\over p}~{<i\ell ^+|T|\bar B>\over <i\ell ^+|T|B>}~.
      \label{rij} \ee
We notice that violations of the $\Delta B=\Delta Q$ rule contribute to
the ratio $\chi _{ij}$ only quadratically. The case of a single channel
say, $i$ (together with $\tilde i$), is obtained by setting $A_j =0$ .

        An instructive but perhaps extreme case arises if we consider
$B\leftrightarrow \bar B$ mixing to be altogether absent and treat the
entire signal of like-sign dileptons to be solely due to the
$\Delta B=- \Delta Q$ transitions. The resulting $\chi _{ij}$ can also be
deduced from the above formulas by taking the limit of vanishing mixing
parameters $x=y=0$ ,
\be \chi _{ij}^{NM}~=~2\left({|A_iA_j|~|\rho _i-\rho _j| \over
          |A_i|^2 + |A_j|^2}\right)^2 ~; \label{chiNM}\ee
the superscript `NM' denotes `no mixing'. Since $|A_i|^2$ is proportional
to the partial width $\Gamma _i$, and hence to the branching fraction
$f_i=\Gamma _i / \Gamma _{total}$, we see that
 \be    |\rho _i~-~\rho _j| = (f_i~+~f_j)\sqrt { {\chi _{ij}^{NM}
           \over 2~ f_i~ f_j} }~.\label{rhodif}\ee

       Experimental data on dilepton events grouped in terms of exclusive
channels are not available. Present data refer to the ratio of inclusive
rates from the ARGUS \cite{ARG87} and CLEO \cite{CLE89,BAR93} groups, and
we take its average value to be  \cite{PDG92}
               \be  \chi _{expt}~= ~0.16\pm 0.04~.\ee
However the signal from the inclusive semileptonic decay (total branching
fraction $\simeq 10 \%$) seems to be arising only from a few channels;
 \[  f(B\rightarrow D^-\ell^+ \nu _\ell)~=~(1.8\pm 0.5)\%~,~~~
                         (\rm {Ref.}~\cite{PDG92})  \]
 \[ f(B\rightarrow D^{*-}\ell^+ \nu _\ell) ~=~ (5.2\pm 0.8)\%~.~~~
                           (\rm {Ref.~}\cite{ARG93})     \]
Therefore in order to get an estimate of the $\rho $'s we may assume that
the contribution to $\chi _{expt}$ is almost entirely due to the above two
decay modes. This allows us to identify them \cite{FN2} with labels $i$
and $j$ in Eq. (\ref{rhodif}), and obtain
     \be   |\rho _i-\rho _j|~=~0.65 \pm 0.10~. \label{numb}\ee
In other words, if we view the like-sign dilepton signal from $\Upsilon
(4S)$ as being purely due to an admixture of the $\Delta B = - \Delta Q$
decay amplitudes, we would require their relative magnitudes to be at
least $\simeq 0.33\pm0.05$ [i.e., half the number appearing in
Eq.(\ref{numb})]. In comparison, the amplitude ratio $\rho (K)$
corresponding to $\Delta S= -\Delta Q$ decays in $K^0 _{\ell 3}$ is known
\cite{PDG92} to be very small, $ \rho (K)=(0.6 \pm 1.8)\% -i(0.3
\pm 2.6)\%~ .$

   An interesting ratio that can be constructed out of same-sign dilepton
events which emerge from two channels $i$ and $j$, is the asymmetry
 \be \alpha _{ij} \equiv \{[ N_{ij}^{++}-N_{ij}^{--}]~/~[N_{ij}^{++}+
   N_{ij}^{--}]\}~. \label{alp}     \ee
Keeping terms upto the bilinear ones in $r_i$ and $r_j$, we see that
\be\alpha _{ij}= {1-|g|^4\over 1+|g|^4}~\left \{~1~+~{4\over 1+|g|^4}
\left [Re~(<r>_{ij})^2 ~-~{|g|^2\over 2}~{1+a\over 1-a}~{\chi _{ij}^{NM}}
                 \right ]\right \}.    \ee
The terms in the square brackets are indeed the correction terms in
expressing the denominator in Eq. (\ref{alp}). Thus a nonzero value
of $\alpha _{ij}$ would require not only mixing ($a \neq 1$) but also
$CP$ violation of the mass-matrix ($|g|\neq 1$). The well-known
inclusive version of $\alpha _{ij}$ given by Okun ${\it {et~ al}}$
\cite{OPZ75} (which assumes the $\Delta B=\Delta Q$ rule) also has these
two requirements. If $\Delta B=-\Delta Q$ transitions were the only source
of the like-sign dilepton events, the asymmetry $\alpha _{ij}$ would vanish
but not the ratio $\chi _{ij}$. Available data on the $CP$-violating
dilepton asymmetry however refer to the inclusive semileptonic channels and
the present experimental value \cite{BAR93} has large uncertainties,
    \[ \alpha ~=~[(1-|g|^4)~/~(1+|g|^4)]~=~(3.1\pm9.6\pm3.2)\%~. \]

   {\it Time-Asymmetries~-~} Consider the time-asymmetry among the events
with same-sign dileptons say, $\ell^+\ell^+$, which result from decays
of neutral $B$'s via either of the two channels, labelled $i$ and $j$ :
 \be {\cal A}_{\ell^+\ell^+}(ij)~=~\{[\nu (ij)-\nu (ji)]~/~
                              [\nu (ij)+\nu (ji)]\}~.\label{app}\ee
The symbol $\nu (ij)$ stands for the number of dilepton events in
which the $\ell^+$ associated with channel $i$ occurs earlier than
the one associated with channel $j$; this is ensured simply by integrating
the rates with respect to the variable $\tau = (t_j-t_i)$  over the range
$0$ to $\infty$. We see that
 \be {\cal A}_{\ell^+\ell^+}(ij)~ =~ {-2 \over 1-a}~[~ax~Im~(r_i-r_j)~+~
                                y~Re~(r_i-r_j)]~;  \label{as1}\ee
similarly we also have
 \be {\cal A}_{\ell^-\ell^-}(ij)= {\nu (\tilde i\tilde j)-\nu (\tilde j
  \tilde i) \over \nu (\tilde i\tilde j)+\nu (\tilde j \tilde i)}={2
  \over 1-a}~[~ ax~ Im~(r_i-r_j)~-~ y~ Re~(r_i-r_j)]~.\label{as2}\ee
Obviously these asymmetries vanish in the case of a single channel
(namely $i=j$) because the initial state $\Upsilon (4S)$ is
antisymmetric under the interchange of the two beons, while the final state
is symmetric (when $i=j$) .

       We next define the time-asymmetry for opposite-sign dilepton events
arising from any of the two exclusive channels $i$ and $j$ ,
  \be  {\cal A}_{\ell^+\ell^-}(ij)~=~
     {[\nu (i\tilde i)+\nu (j\tilde j)+\nu (i\tilde j)+\nu (j\tilde i)]-
      [\nu (\tilde ii)+\nu (\tilde jj)+\nu (\tilde ji)+\nu (\tilde ij)]
 \over [\nu (i\tilde i)+\nu (j\tilde j)+\nu (i\tilde j)+\nu (j\tilde i)]+
   [\nu (\tilde ii)+\nu (\tilde jj)+\nu (\tilde ji)+\nu (\tilde ij)]}~.
                                 \end{equation}
Keeping terms upto the linear ones in $r_i$ and $r_j$ , we obtain this
$CP$-violating asymmetry to be
\be{\cal A}_{\ell^+\ell^-}(ij) ~=~ {4ax \over 1+a}~ Im~<r>_{ij}~,
                    \label{as} \ee
where $<r>_{ij}$ is defined in Eq. (\ref{rij}), and the factor multiplying
it is $\simeq 1.1$ (for the typical values $x\simeq 0.67$ and $y\simeq 0$).
The case of a single channel was reported earlier by one of us
\cite{SAR92}, and generalization to  more than two channels is
straightforward. Notice that this asymmetry, unlike the ones in
Eqs. (\ref{as1},\ref{as2}), does not vanish even if all the $r_i$'s are
equal.  We emphasize that a nonzero value of this asymmetry would
establish the presence of $~\Delta B=-\Delta Q$ transitions in decays of
neutral $B$ mesons, \cite{FN3}.

      A related asymmetry is the one with respect to `channels',
without regard to the leptonic charge; in other words we look at the
difference in rates when the decay channels $j~ {\rm and}~\tilde j$ ,
follow (or precede) the decay channels $\tilde i~{\rm and}~i$:
 \be {\cal A}_{\ell ^+\ell ^-} (i\tilde j + \tilde ij)={[\nu (i \tilde j)+
  \nu (\tilde ij)]-[\nu (\tilde ji)+\nu (j \tilde i)]\over [\nu (i
  \tilde j)+\nu (\tilde ij)]+[\nu (\tilde ji)+ \nu (j\tilde i)]}
                = -~ {2y\over 1+a}~ Re~(r_i-r_j) ~. \label{gvd} \ee
Although this signal is $CP$-conserving, in contrast to Eq. (\ref{as}), it
is unfortunately supressed by the factor $y$ which could be quite small
for the beons.

  {\it Decays to $CP$ Eigenstates~-~} It is now well recognized that at
the forthcoming $B$ factories the main thrust of the experimental effort
would be towards measuring the $CP$-violating asymmetries. Of special
interest are the time-dependent rate asymmetries between $B$ and $\bar B$
decays to specific $CP$ eigenstates $f~(=J/\psi~K_S,~\pi^+\pi^-$, ... ),
as they would enable us to test the $CP$-violating mechanism of the
Standard Model; for a recent review see, e.g.,\cite{NQ92}. How do these
$CP$-violating signals get modified when some of the tag-leptons arise
from $\Delta B=-\Delta Q$ transitions?  This is the item we discuss next.

  Consider the events wherein the tag-lepton is emitted in an exclusive
semileptonic channel $(\tilde i\ell ^-) $ or $(i\ell ^+)$. We express the
asymmetry in terms of the rates summed over the channel
index $i$ as follows :
\be {\cal A}_f~=~ {\Sigma _{i}~{[\nu (\tilde i\ell ^-,f)-\nu (i\ell^+,f)~]
 }\over  \Sigma _{i}~{[\nu (\tilde i\ell ^-,f)+\nu (i\ell^+,f)~]}}~, \ee
where $\nu (i\ell ^+,f)$ represents the number of events arising from
$\Upsilon (4S)$ in which the beon decay to $f$ occurs at any time later
than the related semileptonic beon decay into $(i\ell ^+)$. In the Standard
Model the formula for this asymmetry is
 \be {\cal A}_f^{(SM)}~=~{a~[-2x~ Im~(u_f)+1-|u_f|^2]~-~\Omega~[1+|u_f|^2
 +2y~ Re~(u_f)] \over [1+|u_f|^2+2y~Re~(u_f)]~-~a\Omega [-2x~Im~(u_f)+1
                                                 -|u_f|^2]}~;\ee
  \be  u_f~\equiv g~{<f|~T~|\bar B>\over <f|~T~|B>}~~,~~~\Omega
                          \equiv <B_2|B_1>~=~{1-|g|^2 \over 1+|g|^2}~.\ee
In the limit of $CP$ conservation we have $\Omega =0$ and $u_f=\xi _f$ ,
where $\xi _f$ is the $CP$ eigenvalue, $CP|f>~=\xi _f|f>~=\pm |f>$ ; thus
we see that ${\cal A}_f^{(SM)}=0$ in that limit.

               Going beyond the Standard Model and retaining the
$\Delta B= -\Delta Q$ amplitudes upto first order, i.e., by keeping terms
upto linear ones in the parameter $r_i$ , we obtain
\be {\cal A}_f~=~ \left [~1~+~ {Re~(u_f) +y \over 1+y~Re~(u_f)}~2Re~<r>
\right ]~  {\cal A}_f^{(SM)}~, \label{Af} \ee
where the correction depends on the weighted average of the ratio of the
nonstandard to standard amplitudes,
\be <r>~=~{\Sigma _{i} ~(r_i~|A_i|^2) \over \Sigma _{i}~ |A_i|^2 }~.
                                                \label{RER}\ee
Since ${\cal A}_f^{(SM)}$ is already of first order in $CP$ violation,
and since $<r>$ is expected to be small, Eq. (\ref{Af}) shows that the
$CP$-violating asymmetry ${\cal A}_f$ will be hardly affected by the
presence of $\Delta B=-\Delta Q$ transitions.

             Finally we comment on the interesting asymmetry that provides
a direct measure of the parameter $y$. This needs the same data-base as
the $CP$-violating ${\cal A}_f$ with the difference that the lepton here
serves to fix the `zero' of the time of decay into the $CP$ eigenchannel
$f$. The time-integrated rate for the emission of a lepton of either charge
to emerge ${\it {earlier}}$ than the channel $f$ is
 \be \nu (\ell, f)~\equiv ~\Sigma _i~[\nu (i\ell ^+,f)+
                        \nu (\tilde i\ell ^-,f)]~.         \ee
In a similar way let $\nu (f,\ell )$ determine the rate wherein the
semileptonic decays occur ${\it {after}}$ the decay to $f$; it is obtained
by integrating with respect to the relative time $\tau =(t_f-t_\ell )$
over negative values. With the help of these quantities we formulate
the asymmetry that determines $y$ \cite{DS92},\cite{SAR92}
\be {\cal A}_y={\nu (\ell,f)-\nu (f,\ell )\over \nu (\ell,f)+\nu (f,\ell )}
   =\left [{2Re~(u_f)\over 1+|u_f|^2}\right ]~ \left\{ 1-4(\xi _f -
                            ~Re~ u_f)~Re~<r>\right \}~y ~.  \ee
Thus the value obtainable for $y$ will be hardly modified by the inclusion
of $\Delta B=-\Delta Q$ contributions since the parameter $<r>$ of Eq.
(\ref{RER}) occurs in multiplication with a $CP$-violating effect in the
decay $B \rightarrow f$ .

    In summary, signals for physics beyond the Standard Model could appear
at $B$ factories as nonvanishing time-asymmetries for dilepton events;
Eqs. (\ref{app}-\ref{gvd}). Of particular significance is the asymmetry
${\cal {A}}_{\ell ^+\ell ^-}$ of Eq. (\ref{as}) which is $CP$-violating
and which does not vanish in the limit of $y=0$. On the other hand
$\Delta B=-\Delta Q$ contributions, enter only bilinearly in the
time-integrated dilepton ratios $\chi _{ij}$ and $\alpha _{ij}$ , and
hardly affect the interesting asymmetries which depend on beon decays to
$CP$ eigenstates $f$ .

\end{document}